\documentclass[prl,superscriptaddress,twocolumn,groupedaddress]{revtex4}

\usepackage[highlight]{neel}

\bibpunct{[}{]}{,}{n}{}{}

\graphicspath{{fig/}}

\usepackage[latin1]{inputenc}
\usepackage[T1]{fontenc}

\def\figurename{Fig.}

\usepackage[paperwidth=8.5in, paperheight=11in]{geometry}
\usepackage{graphics} 
\usepackage{times}
\usepackage{bbm}
\usepackage{tabularx}
\usepackage{sistyle}
\usepackage{url}
\usepackage{float}
\usepackage{pdfpages} 

\usepackage{notes2bib}

\newcommand{\CoNi}[2]{$\mathrm{Co}_{#1}\mathrm{Ni}_{#2}$}

\begin{document}
\title{Fast domain wall motion governed by topology and \OE rsted fields in cylindrical magnetic nanowires}

\author{M.~Sch\"{o}bitz}
\email{michael.schobitz@cea.fr}
\affiliation{Univ.\ Grenoble Alpes, CNRS, CEA, Spintec, Grenoble, France}
\affiliation{Friedrich-Alexander Univ.\ Erlangen-N\"{u}rnberg, Inorganic Chemistry, Erlangen, Germany}
\affiliation{Univ.\ Grenoble Alpes, CNRS, Institut N\'{e}el, Grenoble, France}
\author{A.~De Riz}
\affiliation{Univ.\ Grenoble Alpes, CNRS, CEA, Spintec, Grenoble, France}
\author{S.~Martin}
\affiliation{Univ.\ Grenoble Alpes, CNRS, CEA, Spintec, Grenoble, France}
\affiliation{Univ.\ Grenoble Alpes, CNRS, Institut N\'{e}el, Grenoble, France}
\author{S.~Bochmann}
\affiliation{Friedrich-Alexander Univ.\ Erlangen-N\"{u}rnberg, Inorganic Chemistry, Erlangen, Germany}
\author{C.~Thirion}
\affiliation{Univ.\ Grenoble Alpes, CNRS, Institut N\'{e}el, Grenoble, France}
\author{J.~Vogel}
\affiliation{Univ.\ Grenoble Alpes, CNRS, Institut N\'{e}el, Grenoble, France}
\author{M.~Foerster}
\affiliation{Alba Synchrotron Light Facility, CELLS, Barcelona, Spain}
\author{L.~Aballe}
\affiliation{Alba Synchrotron Light Facility, CELLS, Barcelona, Spain}
\author{T.~O.~Mente\c{s}}
\affiliation{Elettra-Sincrotrone Trieste S.C.p.A., Basovizza, Trieste, Italy}
\author{A.~Locatelli}
\affiliation{Elettra-Sincrotrone Trieste S.C.p.A., Basovizza, Trieste, Italy}
\author{F.~Genuzio}
\affiliation{Elettra-Sincrotrone Trieste S.C.p.A., Basovizza, Trieste, Italy}
\author{S.~Le-Denmat}
\affiliation{Univ.\ Grenoble Alpes, CNRS, Institut N\'{e}el, Grenoble, France}
\author{L.~Cagnon}
\affiliation{Univ.\ Grenoble Alpes, CNRS, Institut N\'{e}el, Grenoble, France}
\author{J.~C.~Toussaint}
\affiliation{Univ.\ Grenoble Alpes, CNRS, Institut N\'{e}el, Grenoble, France}
\author{D.~Gusakova}
\affiliation{Univ.\ Grenoble Alpes, CNRS, CEA, Spintec, Grenoble, France}
\author{J.~Bachmann}
\affiliation{Friedrich-Alexander Univ.\ Erlangen-N\"{u}rnberg, Inorganic Chemistry, Erlangen, Germany}
\affiliation{Institute of Chemistry, Saint-Petersburg State Univ., St.\ Petersburg, Russia}
\author{O.~Fruchart}
\affiliation{Univ.\ Grenoble Alpes, CNRS, CEA, Spintec, Grenoble, France}
\email{olivier.fruchart@cea.fr}

\date{\today}

\begin{abstract}

While the usual approach to tailor the behavior of condensed matter and nanosized systems is the choice of material or finite-size or interfacial effects, topology alone may be the key. In the context of the motion of magnetic domain-walls~(DWs), known to suffer from dynamic instabilities with low mobilities, we report unprecedented velocities $>\SI{600}{m/s}$ for DWs driven by spin-transfer torques in cylindrical nanowires made of a standard ferromagnetic material. The reason is the robust stabilization of a DW type with a specific topology by the \OE rsted field associated with the current. This opens the route to the realization of predicted new physics, such as the strong coupling of DWs with spin waves above $>\SI{600}{m/s}$.

\end{abstract}

\maketitle

It is well known that specific properties in condensed-matter and nanosized systems can be obtained by either acting on the electronic structure by selecting an appropriate material composition and crystalline structure, or by making use of finite-size and interfacial effects, strain, gating with an electric field, \etc\cite{bib-NEW2005}. These approaches have proven suitable for tailoring charge transport, optical properties, electric or magnetic polarization, \etc. However, there are limits regarding what can be achieved with materials, or realized with device fabrication. An alternative strategy entails considering a specific topology in order to develop the desired properties of a system, yielding diverse applications such as the design of wide-band-gap photonic crystals\cite{bib-LU2017} and the control of flow of macromolecules\cite{bib-QIN2014}, or novel theoretical methods such as for the description of defects\cite{bib-MER1979}, or intringuing 3D vector-field textures such as hopfions and torons\cite{bib-ACK2017}. As regards magnetism, unusual properties resulting from topological features have been predicted, such as the existence of a domain wall~(DW) in the ground state of a Moebius ring\cite{bib-PYL2015}, or the non-reciprocity of spin waves induced by curvature and boundary conditions in nanotubes\cite{bib-YAN2011b}.

Here, we show that topology plays a critical role in the physics of DW motion in one-dimensional conduits, a prototypical case for magnetization dynamics. For the sake of simplicity of fabrication and monitoring, DW motion under magnetic field or spin-polarized current is usually conducted in planar systems, made of stacked thin films patterned laterally by lithography. In them, DWs are dynamically unstable above a given threshold of field or current (Walker limit), undergoing transformations of their magnetization texture, associated with a drastic drop in their mobility. Ways are being investigated to overcome this limitation through the engineering of microscopic properties. Two major routes are the use of the Dzyaloshinskii-Moriya interaction in order to stabilize the walls\cite{bib-MIR2011,bib-THI2012,bib-RYU2013}, or of natural or synthetic ferrimagnets with vanishing magnetization to decrease the angular momentum in order to switch and boost the precessional frequency\cite{bib-KIM2017c,bib-CAR2018,bib-YAN2015}.

The three-dimensional nature of cylindrical nanowires~(NWs) gives rise to the existence of a DW with a
specific topology, which respects the rotational invariance and circular boundary conditions. It is named the Bloch-point wall~(BPW)\cite{bib-THI2006} and has been experimentally confirmed only recently\cite{bib-BIZ2013,bib-FRU2014}. It was predicted that this wall can circumvent the Walker limit, but field-driven motion experiments disappointingly failed to confirm a topological protection\cite{bib-FRU2019}. Here, we report experimental results on current-induced DW motion in such NWs. We show that although previously disregarded, the \OE rsted field induced by the current plays instead a crucial and valuable role in stabilizing BPWs, contrary to the field-driven case. This allows them to retain their specific topology and thus reach velocities $>\SI{600}{m/s}$ in the absence of Walker breakdown, which is quantitatively consistent with predictions.

\begin{figure}
\centering\includegraphics[width=\linewidth]{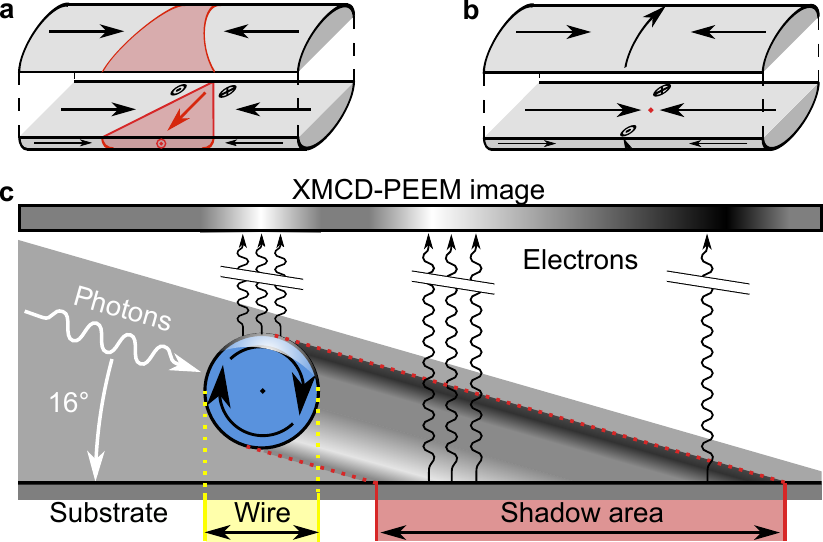}
\caption{Schematic of \textbf{a} a TDW and \textbf{b} a BPW. \textbf{c} Schematic of shadow XMCD-PEEM and the contrast resulting from a BPW.}
\label{fig:figure1}
\end{figure}

DWs with two distinct topologies exist in NWs: the transverse-vortex wall~(TVW) and the BPW\bracketsubfigref{fig:figure1}{a,b}. The former has the same topology as all DW types known in 2D flat strips\cite{bib-FRU2015b}. The latter is found only in NWs and exhibits azimuthal curling of magnetic moments around a Bloch point, a local vanishing of magnetization\cite{bib-FEL1965,bib-DOE1968}. This unique topological feature of NWs is at the origin of the predicted fast speed and stability during magnetic-field or current-driven motion of BPWs. This is easily explained by considering the time derivative of the magnetization vector $\diffdot{\vect m}$ at any point, described by the Landau-Lifshitz-Gilbert equation\cite{bib-THI2005}:

\begin{equation}
\diffdot{\vect m} =  -\gamma_0 \vect m\crossproduct\vect H + \alpha\vect m\crossproduct \diffdot{\vect m} - \left(\vect u \dotproduct \vectNabla \right) \vect m + \beta \vect m \crossproduct \left[ \left(   \vect u \dotproduct \vectNabla   \right) \vect m \right]
\label{eq:LLG}
\end{equation}
with $\gamma_0 = \mu_0 \left| \gamma \right|$, $\gamma$ being the gyromagnetic ratio, $\alpha\ll1$ the Gilbert damping parameter and $\beta$ the non-adiabaticity parameter. $\vect H$, the total effective field, is comprised of applied fields and fields originating from magnetic anisotropy, exchange and dipolar energy. The spin-polarized part of the charge current induces so-called spin-transfer torques, taken into account through $\vect u$, with $|\mathrm{\vect u}| = P\left(j\muB / e\Ms\right)$\cite{bib-THI2005}. $j$ and $P$ are the charge current and its spin-polarization ratio, respectively, $\muB$ is the Bohr magneton, $e$ the elementary charge and $\Ms$ the spontaneous magnetization.

In purely field driven cases, the applied field favors the precession of $\vect m$ around the field direction. In flat strips, for applied fields above a few~mT this causes repeated DW transformations from transverse to vortex walls for in-plane magnetization, and from N\'{e}el to Bloch walls for out-of-plane magnetization. This so-called Walker breakdown\cite{bib-SCH1974} is facilitated by the fact that all these DW configurations share the same topology\cite{bib-BEA2005,bib-MOU2007,bib-HAY2007}. The mobility is high below the Walker threshold field~(scaling with $1/\alpha$) and low above~(scaling with~$\alpha$). The same physics is expected in NWs for the TVW, with the Walker field equal to zero due to the rotational symmetry \cite{bib-THI2006,bib-YAN2011b}. The phenomenology of current-driven cases is similar: the adiabatic term favors motion, the non-adiabatic term favors azimuthal precession~[third and fourth terms in \eqnref{eq:LLG}, respectively], and the DW velocity is expected to be $\approx (\beta/\alpha)u$ below the Walker threshold and $\approx u$ above it\cite{bib-THI2005,bib-MOU2007}, again with a vanishing threshold for TVWs in NWs\cite{bib-YAN2010}.

In contrast to these cases, one expects that magnetization cannot freely precess azimuthally in a BPW, since it would periodically imply a head-on or tail-on configuration along all three axes, with an enormous cost in dipolar energy. Instead, the azimuthal rotation should come to a halt and remain in a state essentially similar to the static one\bracketsubfigref{fig:figure1}b. This implies an absence of Walker breakdown, both under field\cite{bib-THI2006,bib-YAN2011b} and current\cite{bib-WIE2010,bib-OTA2012}, and steady-state motion of the wall. The steady circulation is expected to be clockwise~(CW) with respect to the direction of motion of the DW, while the counterclockwise~(CCW) circulation may undergo a dynamics-induced irreversible switching event to recover the CW circulation and steady state. This picture is valid both for BPWs in wires\cite{bib-THI2006,bib-WIE2010}, and vortex walls\cite{bib-YAN2011b,bib-OTA2012} in thick-walled tubes. Thanks to this locked topology, the mobility of the BPW is expected to remain high under both field and current. Only when a speed around \SI{\approx1000}{m/s} is attained, the speed is predicted to reach a plateau, with new physics expected to occur via interactions with spin waves, known as the spin-Cherenkov effect\cite{bib-YAN2011b}.
However, so far there exists no experimental report of the mobility of any of these walls under neither magnetic field nor current.


Our work is based on magnetically-soft \CoNi{30}{70} wires with diameter \unit[90]{nm}, electroplated in anodized alumina templates\cite{bib-BOC2017}. Following the dissolution of the latter, isolated wires lying on a Si substrate are contacted with pads to allow for the injection of electric current. DWs were monitored with both magnetic force microscopy~(MFM) and X-ray Magnetic Circular Dichroism Photo-Emission Electron Microscopy~(XMCD-PEEM) in the shadow mode \bracketsubfigref{fig:figure1}{c} to reveal the three-dimensional texture of magnetization\cite{bib-KIM2011b,bib-FRU2014,bib-FRU2015c}. While in MFM, sharp ns-long pulses could be sent, in XMCD-PEEM the shape of current pulses was distorted to a minimum width of $\unit[10-15]{\nano\second}$, due to long cabling, UHV feedthroughs and the sample holder contacts. Micromagnetic simulations were carried out with the home-made finite-element code FeeLLGood\cite{bib-FEE}, based on the Landau-Lifshitz-Gilbert equation including spin-transfer torques. See supplementary information for additional details on the methods\footnote{See Supplemental Material [url] for additional details on sample fabrication, MFM and XMCD PEEM imaging, pulse width and DW velocity and error calculations, Joule heating, micromagnetic simulations and DW inertia simulations, which includes Refs.~\cite{bib-KAD1981,bib-IKE1988,bib-NIS1983,bib-TOT2010,bib-ABA2015,bib-FOE2016,bib-WAR1978,bib-OU2007,bib-THI2007,bib-UED2012}}.

\begin{figure}
\centering\includegraphics[width=\linewidth]{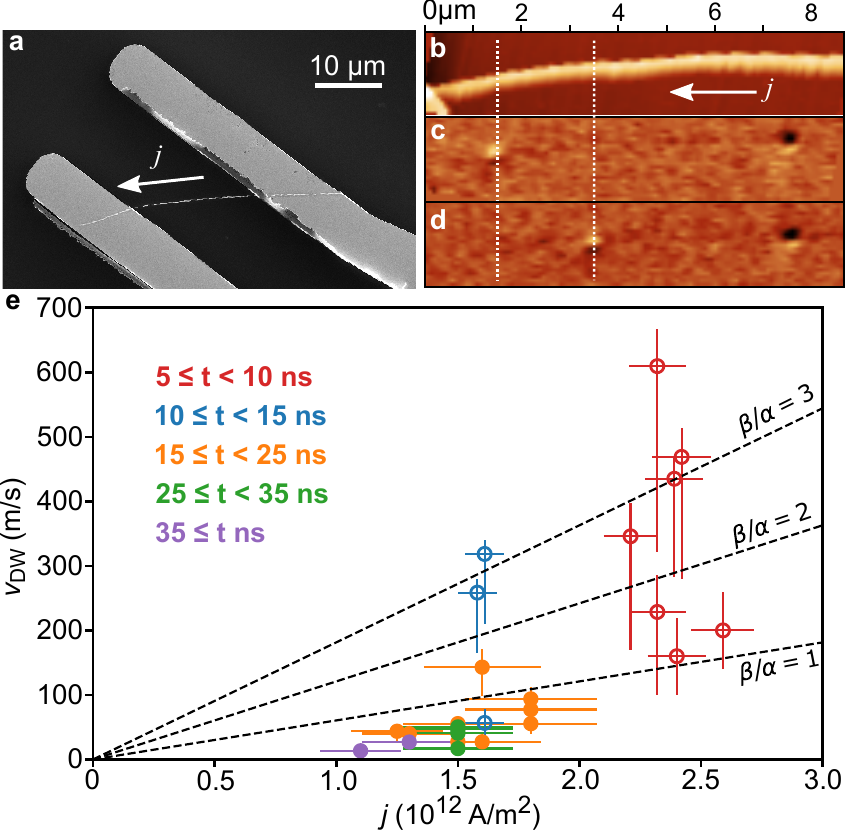}
\caption{\textbf{a} SEM, \textbf{b} AFM and corresponding \textbf{c}, \textbf{d} MFM images of a \SI{90}{nm} diameter Co$_{30}$Ni$_{70}$ NW with Ti/Au electrical contacts. \textbf{c} Initial state, with two DWs \textbf{d} Same wire, after a current pulse with $\SI{2.2\times{{10}^{12}}}{A/m^{2}}$ magnitude and \SI{5.8}{ns} duration.
\textbf{e} Domain wall velocity as a function of applied current density, duration~(see inner caption), monitored with MFM (open circles) and XMCD PEEM (filled circles) from 4 individual NWs. The dashed lines are expectations from the one-dimensional model below the Walker breakdown, for $v=(\beta/\alpha)u$ with $\beta/\alpha=1$, $2$, $3$.}
\label{fig:figure2}
\end{figure}

Domain wall velocities were experimentally investigated primarily with MFM imaging. \subfigref{fig:figure2}{b} shows an atomic force microscopy~(AFM) image of the left hand side of the contacted NW from \subfigref{fig:figure2}a. The corresponding magnetic force microscopy~(MFM) image in \subfigref{fig:figure2}{c} shows the initial magnetic configuration, with two DWs located at $1.2$ and \unit[7.2]{\micro{}m} from the edge of the left contact. By applying a current pulse of duration \SI{5.8}{ns} and amplitude $\SI{2.2\times{{10}^{12}}}{A/m^{2}}$, the left hand DW moved over a distance of $\approx$~\SI{2}{\micro{}m}\bracketsubfigref{fig:figure2}{d}, corresponding to an average velocity of $\approx$~\SI{350}{m/s}. However, the right hand DW remains pinned, highlighting a common and key issue for inferring DW velocities from motion distances: pinning on geometrical or microstructural defects hampers DW motion\cite{bib-FRU2016c}. Depinning not only requires a current density above a critical value $j_{\mathrm{dp}}$, but re-pinning can also occur at another location with a deeper energy well, while the current pulse is still being applied. This results in DW propagation with an effective time span possibly much shorter than the nominal pulse duration. Consequently, the values for DW velocity converted from motion distance and nominal pulse length are a lower bound of an unknown higher velocity (see supplementary material for a quantitative discussion). Furthermore, with such large current densities the effect of Joule heating may not be neglected. However, measurements of the NW resistance during the pulse showed that the samples never exceeded the Curie temperature (see supplementary material) and that the results described herein are not caused by thermal activation.

\subfigref{fig:figure2}{e} (open circles) shows the discussed lower bound for DW velocity, as a function of applied current density, inferred from a multitude of MFM images before and after pulses with durations ranging from $5$ to \SI{15}{ns}. Consistent with the expected occurrence of re-pinning, lower velocities are inferred from longer pulse durations. Still, DW velocities up to $>\SI{600}{m/s}$ were observed for applied current densities $\approx\SI{2.4\times{{10}^{12}}}{A/m^{2}}$. This sets a five-fold record for purely spin-transfer torque motion of DWs in a standard ferromagnetic material, \ie, with large magnetization, with reported values hardly exceeding $\SI{100}{\meter\per\second}$\cite{bib-BOU2011}. Similar or higher speeds have been measured recently, however in low-magnetization ferrimagnets, thereby enhancing the efficiency of spin-transfer torque\cite{bib-GUS2019}. Here, it is the topology of the wall that enhances the DW speed, not a special material. Similarly, these DW velocity measurements are not enhanced by DW inertia, since simulations showed that this effect will only come into play in sub-nanosecond pulse experiments (see supplementary material). The black dotted lines in \subfigref{fig:figure2}{e} act as a guide to the eye through the speed predicted by the one-dimensional model below the Walker breakdown $v=(\beta/\alpha)u$, for three different ratios of $\beta/\alpha$: 1, 2 and~3~(for \CoNi{30}{70} $\Ms=\SI{0.67}{\mega{}A/m^{2}}$, $P\approx0.7$, resulting in $u\approx\SI{60.4}{m/s}$ per $\SI{10^{12}}{A/m^{2}}$). This is not intended as precise modelling, but rather to show that the experiments are clearly not compatible with $v=u$, supporting the absence of Walker breakdown for the BPW. Instead a value of $\beta/\alpha\gtrapprox3$ is inferred.  Note, however, that the adverse effects of DW pinning reappear in the form of a threshold current density $j_{\mathrm{dp}}\approx\SI{1.2\times{{10}^{12}}}{A/m^{2}}$ required to set any DW in motion. Even above this value, DW motion was not fully reproducible, with some pinning sites associated with a larger $j_{\mathrm{dp}}$.



\begin{figure}
\centering\includegraphics[width=\linewidth]{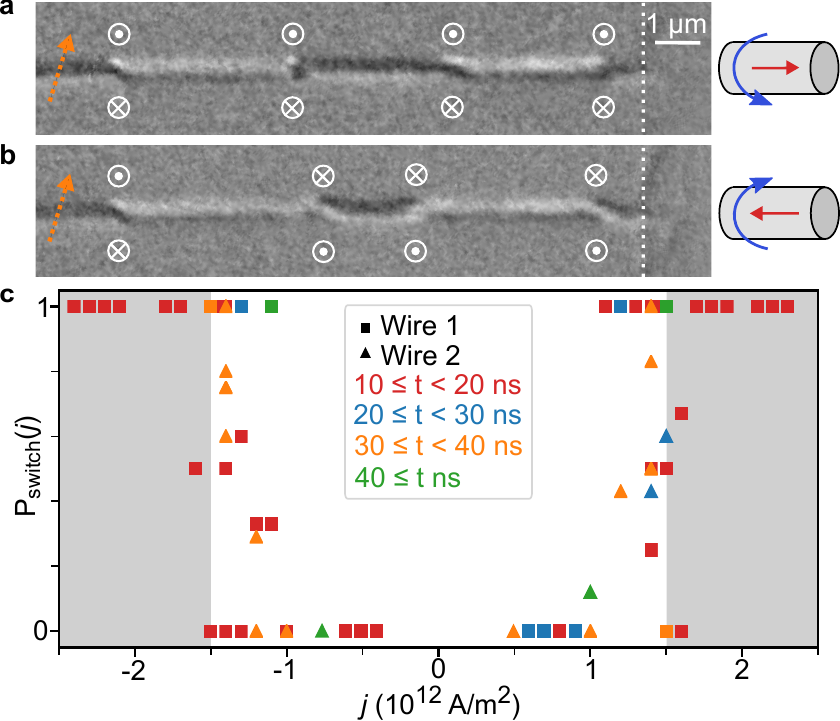}
\caption{\textbf{a}, \textbf{b} Consecutive XMCD-PEEM images of a NW with a tilted x-ray beam (orange arrow). The azimuthal circulation of the four BPWs seen in the NW shadow is indicated by the white arrows, consistent with the \OE rsted field of the previously applied current (blue and red arrows in the right hand schematic, respectively). From \textbf{a} to \textbf{b}, a \SI{15}{ns} and $\SI{1.4\times{{10}^{12}}}{A/m^{2}}$ current pulse switches $75$\% of BPWs. DW displacement from \textbf{a} to \textbf{b} cannot be discussed as directly resulting from spin-transfer torque, and the density of current lies below the threshold for free motion
\textbf{c} BPW switching probability as a function of $j$ for two different wire samples (squares and triangles). Pulse durations are categorized and color coded, see included labels. The grey region indicates the current density required for switching in simulations.}
\label{fig:figure3}
\end{figure}

To link unambiguously the measured velocity with theory, the DW type must be identified. For this purpose, we employed shadow XMCD-PEEM and imaged NWs before and after injecting a given current pulse (\subfigref{fig:figure3}{a,b}, and full symbols in \figref{fig:figure2}e). Note that the values for speed are lower than those measured with MFM, as expected for less sharp pulse shapes with consequentially larger width. Returning to the DW type, the first striking fact is the following: from hundreds of DWs imaged after current injection, all were of the BPW type. These unambiguously appear as a symmetric bipolar contrast in the shadow\cite{bib-FRU2014}, corresponding to azimuthal rotation of magnetization as on \subfigref{fig:figure3}{a-b}. This sharply contrasts with all our previous observations of NWs, imaged in the as-prepared state or following a pulse of magnetic field, for which both TVWs and BPWs had been found in sizeable amounts\cite{bib-FRU2014,bib-FRU2019}. The second striking fact is that the sign of the BPW circulation is deterministically linked to the sign of the latest current pulse, provided that its magnitude is above a rather well-defined threshold which, as shown in \subfigref{fig:figure3}{c}, lies around $\SI{1.4\times{{10}^{12}}}{A/m^{2}}$. In contrast with a one-time Walker event discussed previously, this holds true irrespective of whether or not the wall has moved under the stimulus of the current pulse, and is independent of the pulse duration at the probed timescales. We hypothesize that these two facts are related to the \OE rsted field associated with the longitudinal electric current, its azimuthal direction favoring the BPW with a given circulation. Indeed, for a uniform current density~$j$, the \OE rsted field is $H_\mathrm{\OE} = jr/2$ at distance $r$ from the NW axis. For the present NWs with radius $R=$ \SI{45}{nm} and $j=\SI{1\times{{10}^{12}}}{A/m^{2}}$ this translates to \SI{28}{mT} at the NW surface, which is a significant value.

\begin{figure}
\centering\includegraphics[width=\linewidth]{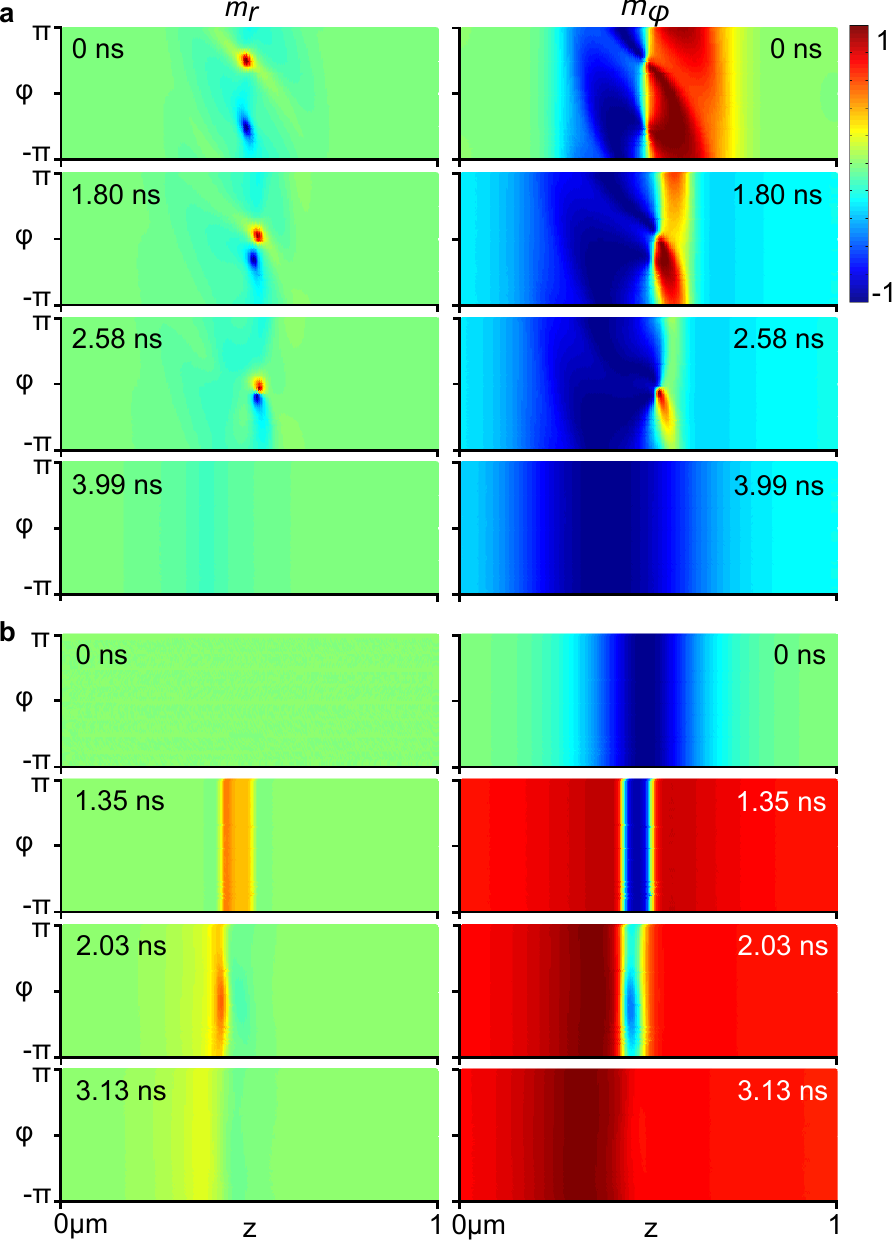}
\caption{DW transformations by the \OE rsted field in micromagnetic simulations for \textbf{a} TVW to BPW, with $j=\SI{0.4\times{{10}^{12}}}{A/m^{2}}$, and \textbf{b} BPW circulation reversal, with $j=\SI{-1.8\times{{10}^{12}}}{A/m^{2}}$. Left and right are color maps of the radial and azimuthal magnetization components, $m_r$ and $m_\varphi$, respectively, over time on the unrolled surface of a \SI{90}{\nano{}m} diameter, \SI{1}{\micro{}m}-long NW with $\alpha = 1$.}
\label{fig:figure4}
\end{figure}

In order to support this claim, we conducted micromagnetic simulations including the \OE rsted field, which had not been considered in previous works. Starting from a DW at rest with $R=\SI{45}{nm}$, we used $\alpha=1$ to avoid ringing effects and obtain a quasistatic picture, suitable to describe the PEEM experiments, for which the pulse rise time is several nanoseconds. We evidenced that while the added effect of spin-transfer torques may alter the transformation mechanisms, it is of second-order compared to the \OE rsted field and considering or disregarding these torques does not quantitatively impact switching. Accordingly, below we present only results disregarding these torques. Within the domains the peripheral magnetization tends to curl around the axis, while it remains longitudinal on the NW axis. We first consider TVWs as the initial state and find that these transform into BPWs with CW circulations with respect to the current direction, if the current density exceeds $\SI{0.4\times{{10}^{12}}}{A/m^{2}}$. The underlying process is illustrated on \subfigref{fig:figure4}{a}, displaying maps of the radial and azimuthal magnetization components, $m_r$ and $m_\varphi$, respectively, on the unrolled surface of a NW as a function of time. These highlight the locations of the inward and outward flux of magnetization through the surface, signature of a TVW\cite{bib-FRU2015b}. While these local configurations are initially diametrically opposite, they approach each other until they eventually merge, expelling the transverse core of the wall from the NW. This is associated with the nucleation of a Bloch point at the NW surface, which later on drifts towards the NW axis, ending up in a BPW. This process is similar to the dynamical transformation of a TVW into a BPW upon motion under a longitudinal magnetic field\cite{bib-FRU2019}, and explains the absence of TVWs in our measurements, for which the applied current densities were always larger than $\SI{0.4\times{{10}^{12}}}{A/m^{2}}$. In order to understand the unique circulation observed, we now consider a BPW as the initial state. BPWs with a circulation matching that of the \OE rsted field do not change qualitatively, only their width increases during the pulse. On the contrary, BPWs shrink if their initial circulation is CCW, \ie opposite to the \OE rsted field. For $j \leq \SI{1.5\times{{10}^{12}}}{A/m^{2}}$ the CCW BPW reaches a narrow yet stable state, and recovers its initial state after the pulse. Beyond this value the circulation switches through a transient radial orientation of magnetization\bracketsubfigref{fig:figure4}{b}. After the switching of circulation, the BPW expands and reaches a stable CW state. The value of the critical current density required for circulation switching is in quantitative agreement with the experimental one~(\subfigref{fig:figure3}{c}, $\approx\SI{1.4\times{{10}^{12}}}{A/m^{2}}$), although the simulation does not incorporate thermal activation and considers $\alpha=1$. This suggests that the switching process is robust and intrinsic, in agreement with the narrow experimental distribution of critical current. In our simulations the time required for switching is \SI{<10}{\nano\second}, though switching times an order of magnitude faster are expected for realistic values of $\alpha<0.1$, which explains why no dependence on the pulse width was observed in the experiments, where all pulse widths were above \SI{5}{\nano\second}.

In experiments where the DW type was visible, DW motion events were observed for applied current densities larger than the critical current density required for the circulation switching event. Thus, in these the circulation is always CCW with respect to the propagation direction, \ie CW with respect to the current direction, because the charge of electrons is negative. Remarkably, this sense of circulation is opposite to the situation expected when neglecting the \OE rsted field, which would select the CW circulation with respect to the propagation direction, as dictated by the chirality of the LLG equation\cite{bib-THI2006,bib-WIE2010,bib-OTA2012}. There must therefore be a competition for the circulation sense and for the case of \SI{90}{\nano{}m} diameter NWs, the \OE rsted field dominates. Despite this, we find that the BPW motion still follows $v\approx(\beta/\alpha)u$ whether or not the \OE rsted field is considered. Notice that the $\beta$ parameter is expected to depend on the DW width, however for widths much smaller than the ones studied here\cite{bib-STU2016}. The predictions of high mobility and possibly spin-Cherenkov effect are thus probably not put into question.

Surprisingly, the \OE rsted field was previously only considered in a single report for NWs of square cross-section\cite{bib-AUR2013}. No qualitative impact was found, likely because a NW side of at most $\SI{48}{\nano\meter}$ was considered, and a simple analytical model describing magnetization in the domain and balancing Zeeman \OE rsted energy with exchange energy shows that the impact of the \OE rsted field scales very rapidly as $R^3$. This is also accurately confirmed by simulations. The situation closest to the present case is the report of flat strips made of spin-valve asymmetric stacks\cite{bib-FRU2011f}. Such strips can be viewed as the unrolled surface of a wire, the curling of the BPW translating into a transverse wall, which tends to be stabilized during motion due to the \OE rsted field.


To conclude, we have shown experimentally and by simulation that the \OE rsted field generated by the spin-polarized current flowing through a cylindrical NW has a crucial impact on DW dynamics, while it had been disregarded so far. This \OE rsted field robustly stabilizes BPWs, in contrast with the field-driven case\cite{bib-FRU2019}. This stabilization allows for the key features predicted for their specific topology to apply\cite{bib-THI2006,bib-WIE2010,bib-OTA2012}: we evidenced DW velocities in excess of \SI{600}{m/s} confirming the absence of Walker breakdown\cite{bib-YAN2011b,bib-HER2016} and setting a five-fold record for spin-transfer-torque-driven DW motion in large magnetization ferromagnets\cite{bib-BOU2011}. This suggests that the experimental realization of further novel physics is at hand, such as the predicted spin-Cherenkov effect with strong coupling of DWs with spin waves.

\section{Acknowledgments}

MS acknowledges a grant from the Laboratoire d'excellence LANEF in Grenoble (ANR-10-LABX-51-01). The project received financial support from the French National Research Agency (Grant No. JCJC MATEMAC-3D). This work was partly supported by the French RENATECH network, and by the Nanofab platform (Institut N\'{e}el), whose team is greatly acknowledged for technical support. We thank Jordi Prat for his technical support at the ALBA Circe beamline and Olivier Boulle for useful discussions.

\bibliographystyle{apsrev4-1}
\bibliography{Fruche7}

\end{document}